\documentclass[12pt]{article}
\usepackage[english]{babel}
\usepackage{color}
\usepackage{graphicx}
\usepackage{amsmath}
\usepackage{amsthm}
\usepackage{amssymb}
\usepackage[hyperindex,plainpages=false]{hyperref}
\usepackage[left=1.1in,right=1.1in,bottom=1in,top=1in,a4paper]{geometry}

\title{The Gardner method for symmetries}
\author{Alexander G. Rasin  \\
Department of Computer Science and Mathematics,\\ 
Ariel University, Ariel 40700, Israel \\
{E-mail: rasin@ariel.ac.il}\and  Jeremy Schiff \\
Department of Mathematics,\\
Bar-Ilan University, Ramat Gan, 52900, Israel \\
{E-mail: schiff@math.biu.ac.il}}

\begin{document}
\maketitle
\begin{abstract}{
The Gardner method, traditionally used to generate conservation laws
of integrable equations,
is generalized to generate symmetries. The method is demonstrated for
the KdV, Camassa-Holm and Sine-Gordon
equations. The method involves identifying a generating symmetry which depends
upon a parameter;
expansion of this symmetry in a (formal) power series in the
parameter then gives
the usual infinite hierarchy of symmetries. We show that the obtained
symmetries commute, discuss the
relation of the Gardner method with Lenard recursion (both for
symmetries and
conservation laws), and also the connection between the symmetries of
continuous integrable
equations and their discrete analogs.}
\end{abstract}

\section{Introduction}
More than forty years ago Miura discovered the so-called Miura map \cite{Miu0}.
In a subsequent paper Miura, Gardner and Kruskal showed how to
use the Miura map for the construction of an infinite hierarchy of conservation laws
for the Korteweg-de Vries (KdV) equation \cite{Miu1}. For brevity we call this
technique the Gardner method. Other techniques for the construction of
conservation laws have appeared since then, for example Lenard recursion
\cite{GGK,Lax2,Lax3,Mag0,Ol0},
the Gelfand-Dickey method \cite{Dic0,GDi1,Wil1} and the symmetry method \cite{FF0,Ol0}.
Nevertheless the Gardner method is considered to be the simplest and most elegant method for the
construction of conservation laws. It has been applied to other integrable equations, for instance to the Camassa-Holm(CH) equation \cite{FS1}.

An analog of the Gardner method for discrete equations appeared recently \cite{Ra1, RS1}.
An infinite number of conservation laws were constructed for discrete KdV
(dKdV) and all the ABS equations \cite{ABS} using B\"acklund transformations (BTs ---
note the Miura map is one of the  defining equations for the BT of KdV). Moreover in
\cite{Ra1} it was shown that it is also possible to use the Gardner method  to construct the
infinite hierarchy of symmetries for dKdV. This result raises the question whether it is
possible to do the same for continuous equations.

The theory of symmetries for continuous equations is well developed. There is a direct
method which allows computation of all symmetries of a given order for a given equation.
There are numerous methods to generate the infinite hierarchies of symmetries of
integrable equations such as KdV, for example, the mastersymmetry method \cite{Fu0},
Lenard recursion \cite{Ol1,Ol0,PS0}  and Lax operator methods \cite{GDi1,Lax2,Wil5}. The method we will present in this paper has much in common with the method of the resolvent, described, for KdV, in section 3.7 of \cite{Dic0}.

The goal of this paper is to present the Gardner method for symmetries of
continuous partial differential equations (PDE). It is a little more subtle than the Gardner method for
conservation laws, but it is still simple and elegant.
The structure of this paper is as follows: In section 2 the
Gardner method for symmetries is presented for KdV, CH and sine-Gordon (SG) equations.
In section 3 we prove that the obtained
symmetries commute. In section 4 we show the connection of the Gardner method and Lenard recursion
(both for symmetries and for conservation laws).
In section 5 the connection between the symmetries of continuous and
discrete equations is described. Section 6 contains some concluding comments
and questions for further study.

\section{Gardner method for symmetries}
\subsection{KdV}
The potential KdV (pKdV) and KdV equations have the form
\begin{eqnarray}
u_t-\textstyle{\frac{3}{2}}u_x^2-\textstyle{\frac{1}{4}}u_{xxx}&=&0\ , \label{eq1} \\
\phi_t-3\phi\phi_x-\textstyle{\frac{1}{4}}\phi_{xxx} &=& 0 \label{eq}
\end{eqnarray}
related by $\phi=u_x$. The standard B\"acklund transformation for (p)KdV \cite{DrJo}
can be written 
$u\rightarrow u_\alpha=u+v_\alpha$ where $v_\alpha$ satisfies
\begin{eqnarray}
v_{\alpha,x} &=&\alpha-2\phi-v_{\alpha}^2 \ ,  \label{BT10}\\
v_{\alpha,t} &=&-\textstyle{\frac12} \phi_{xx}
    + (\alpha+\phi)(\alpha -2\phi-v_\alpha^2) + \phi_{x}v_\alpha\ .  \label{BT11}
\end{eqnarray}
The index $\alpha$ indicates that $v_{\alpha}$ is the solution of (\ref{BT10})-(\ref{BT11})  
with parameter $\alpha$. The system (\ref{BT10})-(\ref{BT11}) 
is consistent if and only if $\phi$ satisfies the KdV
equation (\ref{eq}) and has a one-parameter family  of  solutions.
There is an algebraic way to describe the action of repeated B\"acklund transformations
with different parameters \cite{DrJo}. This relation is called the nonlinear
superposition principle and has the form
\begin{equation}
u_{\alpha,\beta}=u+\frac{\beta-\alpha}{u_{\beta}-u_{\alpha}}.\label{eq2}
\end{equation}
Here $u_{\alpha}=u+v_{\alpha}$ is obtained by application of the BT with parameter $\alpha$,
$u_{\beta}=u+v_{\beta}$ is from application of the BT with parameter $\beta$,
and $u_{\alpha,\beta}$ is the result of application of both, in either order, as they
commute. Writing $\beta=\epsilon+\alpha$, relation (\ref{eq2}) can be rewritten
\begin{equation}
u_{\alpha,\alpha+\epsilon}
=u+\frac{\epsilon}{u_{\alpha+\epsilon}-u_{\alpha}}
=u+\frac{\epsilon}{v_{\alpha+\epsilon}-v_{\alpha}}
.\label{eq30}
\end{equation}
For small $\epsilon$ we can expand $v_{\alpha+\epsilon}$ as a power series in $\epsilon$.
To lowest order, $v_{\alpha+\epsilon}$ satisfies the same system of differential equations
as $v_\alpha$. But this does  {\em not} mean that to lowest order it is the same as
$v_\alpha$, since, as we have explained, the BT has a one-parameter family of solutions.
Relation (\ref{eq30}) can thus be rewritten in the form
\begin{equation}
u_{\alpha,\alpha+\epsilon}
=u+\frac{\epsilon}{v_{\alpha}^{(1)}-v_{\alpha}^{(2)}}+O(\epsilon^2)\label{eq40}.
\end{equation}
where $v_{\alpha}^{(1)}$ and $v_{\alpha}^{(2)}$ are distinct solutions of
(\ref{BT10})-(\ref{BT11}) for the same parameter value $\alpha$.

Relation (\ref{eq40}) describes an infinitesimal continuous transformation of
$u$ which is also a solution of pKdV. In other words, this is a symmetry for pKdV. 
We denote it  $X(\alpha) = Q(\alpha) \frac{\partial}{\partial u}$ where
\begin{equation}
Q(\alpha)=\frac{1}{v_{\alpha}^{(1)}-v_{\alpha}^{(2)}}.\label{sym1}
\end{equation}
We call $X(\alpha)$ \textit{the generating symmetry}.
Note that while the explanation we have given above for why $X(\alpha)$ is a symmetry 
is perfectly rigorous, there is a more direct technical proof: If $Q$ is defined as in
(\ref{sym1}) and $v_{\alpha}^{(1)}$ and $v_{\alpha}^{(2)}$ are both solutions of
(\ref{BT10})-(\ref{BT11}) then it is a technical exercise to check that
$Q$ satisfies the equation
\begin{equation}
Q_t - 3\phi Q_x - \textstyle{\frac{1}{4}}Q_{xxx}=0\ ,
\end{equation}
this being the defining equation for infinitesimal symmetries of pKdV (\ref{eq1}).

The next thing to do is to
observe that if we could solve (\ref{BT10})-(\ref{BT11})
to write $v_{\alpha}$ as a function of
$\phi$ (or $u$) and (a finite number of) its derivatives, as well as a constant of
integration,  then we could rewrite (\ref{sym1})
in terms of $\phi$. This cannot be done explicitly. However it is possible to write a
(formal) asymptotic series solution of (\ref{BT10})-(\ref{BT11}) for large $|\alpha|$, 
as a series in decreasing powers of $\alpha^{1/2}$. This takes the form
\begin{equation}
v_{\alpha}= \alpha^{1/2}
    - \frac{\phi}{\alpha^{1/2}}
    + \frac{\phi_{x}}{2 \alpha}
    - \frac{\phi_{xx}+2\phi^2}{4\alpha^{3/2}}
    + \frac{\phi_{xxx}+8\phi \phi_{x}}{8\alpha^{2}}
   - \frac{\phi_{xxxx}+8\phi^3+10\phi_{x}^2+12\phi\phi_{xx}}{16\alpha^{5/2}}
    + O\left(\alpha^{-3}\right)
\end{equation}
or
\begin{equation}
v_{\alpha}=\alpha^{\frac12}+\sum_{i=1}^{\infty}\frac{c_i}{\alpha^{\frac{i}{2}}}
\end{equation}
where
\begin{equation}
c_1=-\phi\ ,
\qquad
c_2 =\frac{\phi_{x}}{2}\ ,
\qquad
c_{n+1} = -\frac12\left( c_{n,x} + \sum_{i=1}^{n-1} c_i c_{n-i} \right)\ ,
\qquad n=2,3... \ .
\label{thecs}\end{equation}
At first glance this seems insufficient for our purpose, as it only gives
a single solution of equations (\ref{BT10})-(\ref{BT11}), with the prescribed asymptotic
behavior for large $|\alpha|$. But there evidently is a second solution in
which $\alpha^{1/2}$ is replaced by $-\alpha^{1/2}$. Thus we can take
\begin{equation}
v_{\alpha}^{(1)} = \alpha^{\frac12}+\sum_{i=1}^{\infty}\frac{ c_i}{\alpha^{\frac{i}{2}}} \ , \qquad
v_{\alpha}^{(2)} = -\alpha^{\frac12}+\sum_{i=1}^{\infty}\frac{(-1)^ic_i}{\alpha^{\frac{i}{2}}} \ ,
\label{exps}\end{equation}
with the $c_i$ defined as before to obtain
\begin{eqnarray}
Q(\alpha) &=&
\frac{1}{2\alpha^{1/2}}
\frac1{\left(
1 + \sum_{j=0}^\infty \frac{ c_{2j+1}}{\alpha^{j+1}}
\right)} \label{f1Q}\\
&=&
\frac{1}{2\alpha^{1/2}}\left(
1
+ \frac{\phi}{\alpha}
+\frac{\phi_{xx}+6\phi^2}{4\alpha^2}
+\frac{\phi_{xxxx}+40\phi^3+10\phi_{x}^2+20\phi\phi_{xx}}{16\alpha^3}
+ \ldots
\right)\ .  \nonumber
\end{eqnarray}
This expansion gives the infinite hierarchy of symmetries of pKdV.
The first few  symmetries take the form
\begin{eqnarray*}
X_{0}&=&\frac{\partial}{\partial u}\ , \\
X_{1}&=&u_x\frac{\partial}{\partial u}\ , \\
X_{2}&=&\left(u_{xxx}+6u_x^2\right)\frac{\partial}{\partial u}
      ~= ~ 4u_t \frac{\partial}{\partial u}\ , \\
X_{3}&=&\left(u_{xxxxx}+40u_x^3+10u_{xx}^2+20u_xu_{xxx}\right)
         \frac{\partial}{\partial u} \ .
\end{eqnarray*}
It is straightforward to verify that the $j$'th symmetry (in the above numbering) 
depends on $x$-derivatives of $u$ of order up to $2j-1$, and that the coefficient of the highest
derivative is nonzero, thus guaranteeing nontriviality.

The corresponding symmetry for KdV is
\begin{equation}
Y(\alpha) = Q(\alpha)_x \frac{\partial}{\partial\phi} \ .
\end{equation}
(Note that in general it is nontrivial that a local symmetry of pKdV should give
a local symmetry of KdV, as $u$ is nonlocal in $\phi$, but since all the
symmetries for pKdV we are considering are determined by $\phi$ this is not an issue
here.) Using (\ref{sym1}),(\ref{BT10}) and (\ref{exps}) we have
\begin{equation}
Q(\alpha)_x
= -\frac{v^{(1)}_{\alpha,x}-v^{(2)}_{\alpha,x}}{(v_{\alpha}^{(1)}-v_{\alpha}^{(2)})^2}
= \frac{v_{\alpha}^{(1)}+v_{\alpha}^{(2)}}{v_{\alpha}^{(1)}-v_{\alpha}^{(2)}}
=
\frac{1}{\alpha^{3/2}}
\frac{\left(  \sum_{j=0}^\infty \frac{ c_{2j+2}}{\alpha^{j}}  \right)}
{\left( 1 + \sum_{j=0}^\infty \frac{ c_{2j+1}}{\alpha^{j+1}}  \right)} \ .
\label{f2Q}
\end{equation}
Expanding in powers of $\frac1{\alpha}$, or just differentiating the relevant
formulas for pKdV, gives the first few symmetries for KdV
\begin{eqnarray*}
Y_{1}&=&\phi_x\frac{\partial}{\partial \phi}\ , \\
Y_{2}&=&\left(\phi_{xxx}+12\phi\phi_x\right)\frac{\partial}{\partial \phi}
      ~= ~ 4\phi_t \frac{\partial}{\partial \phi}\ , \\
Y_{3}&=&\left(\phi_{xxxxx}+120\phi^2\phi_x +40\phi_{x}\phi_{xx}+20\phi\phi_{xxx}\right)
        \frac{\partial}{\partial \phi} \ .
\end{eqnarray*}
Thus we can generate explicit formulas for the infinite hierarchy of symmetries of pKdV
or KdV using (\ref{thecs}) to find the $c_i$ and then expanding (\ref{f1Q})
for pKdV or (\ref{f2Q}) for KdV in inverse powers of $\alpha$.

The function $Q(\alpha)$ that has emerged as a generating function for symmetries of KdV
can be identified with the ``resolvent'' of KdV (see \cite{Dic0} section
3.7). Writing $v_\alpha=\frac
{\psi_{\alpha,x}}{\psi_\alpha}$, equation (\ref{BT10}) becomes the
Schr\"odinger equation
$$ \psi_{\alpha, xx} = (\alpha - 2\phi)\psi_{\alpha}\ . $$
Using (\ref{sym1}) and the fact that the Wronskian of two solutions of the
Schr\"odinger equation
is a constant, we find that $Q(\alpha)$ can be identified with the product of
two solutions of the Schr\"odinger equation, which is the resolvent.
However, the proof
we have given above that $X(\alpha)$ is a symmetry of pKdV is new, and
as we shall see,
this new proof allows generalization to other equations, as well as a
simpler proof of properties such as commutativity.

A natural question to ask at this stage is whether the generating symmetry 
$X(\alpha)$ includes more information than the standard infinite hierarchy of 
commuting symmetries of KdV (commutativity will be proved in our approach in section 
3). So far we have only considered the consequences of taking 
$v_{\alpha}^{(1)}$ and $v_{\alpha}^{(2)}$ in (\ref{sym1}) to be the specific solutions of 
(\ref{BT10})-(\ref{BT11}) given by  (\ref{exps}). Using the fact that if a single 
solution of the Riccati equation (\ref{BT10}) is known then it is possible to find
the general solution by quadratures, it is possible to rewrite (\ref{sym1}) in 
the form 
\begin{equation}
Q(\alpha)=  \int_{x_0}^x  e^{2\int_y^x v_{\alpha}^{(1)}(z)dz}  dy + C e^{2\int_{x_0}^x v_{\alpha}^{(1)}(z)dz} 
\end{equation}
where $C$ is an arbitrary constant. Since a linear combination of symmetries is a 
symmetry, both terms on the RHS must individually be generators of symmetries. The 
nonlocal, noncommuting symmetries associated with the second term were considered recently in 
\cite{LHC}. We leave a fuller study of the content of the generating symmetry 
$X(\alpha)$ to a further publication. 

\subsection{CH}
In \cite{FS1} the conserved quantities of CH were derived from those of the associated Camassa-Holm (ACH), introduced in \cite{Je1} and studied further in \cite{hone1999exact,Hon0,Iva0,Rey0}. The ACH equation has the form
\begin{equation}
p_t=p^2f_x,~~~f=\frac{p}{4}\left(\frac{p_t}{p}\right)_x-\frac{p^2}{2}.\label{ACH}\end{equation}
In \cite{Je1} a BT for ACH was given in the form $p\rightarrow p_{\alpha}=p-s_{\alpha,x}$ where $s_\alpha$ satisfies
\begin{align}
s_{\alpha,x}=&-\frac{s_\alpha^2}{p\alpha}+\frac{\alpha}{p}+p,\label{tr1}\\
s_{\alpha,t}=&-s_\alpha^2+\frac{p_t}{p}s_\alpha+\alpha(\alpha-2f).\label{tr2}
\end{align}
The superposition principle for ACH is
\begin{equation}
p_{\alpha,\beta}=p-\left(\frac{(\alpha-\beta)(\alpha\beta-s_{\alpha}s_{\beta})}{\beta s_{\alpha}-\alpha s_{\beta}}\right)_x.\label{BTACH}
\end{equation}
Here $p_{\alpha}=p-s_{\alpha,x}$ is obtained by application of the BT with parameter $\alpha$,
$p_{\beta}=p-s_{\beta,x}$ is from application of the BT with parameter $\beta$,
and $p_{\alpha,\beta}$ is the result of application of both.
Writing $\beta=\epsilon+\alpha$, and  expanding (\ref{BTACH}) around $\epsilon=0$ we obtain
\begin{equation}
p_{\alpha,\alpha+\epsilon}=p-\epsilon\left(\frac{\alpha^2-s_{\alpha}^{(1)}s_{\alpha}^{(2)}}{s_{\alpha}^{(1)}-s_{\alpha}^{(2)}}\right)_x+O(\epsilon^2)\label{exACH}
\end{equation}
where $s_{\alpha}^{(1)}$ and $s_{\alpha}^{(2)}$ are distinct solutions of (\ref{tr1},\ref{tr2}) for the same parameter value $\alpha$.
Relation (\ref{exACH}) describes an infinitesimal continuous transformation of $p$ which is also
a solution of ACH. We denote it  $X(\alpha) = Q(\alpha) \frac{\partial}{\partial p}$ where
\begin{equation}
Q(\alpha)=\left(\frac{\alpha^2-s_{\alpha}^{(1)}s_{\alpha}^{(2)}}{s_{\alpha}^{(1)}-s_{\alpha}^{(2)}}\right)_x=\frac{p(s_{\alpha}^{(1)}+s_{\alpha}^{(2)})}{s_{\alpha}^{(1)}-s_{\alpha}^{(2)}}.\label{symACH}
\end{equation}
Using (\ref{ACH},\ref{tr1},\ref{tr2}) it can be checked that the coefficient of $\frac{\partial}{\partial f}$ in the prolongation of $X(\alpha)$ is
\begin{equation}
\tilde{Q}=\frac{\alpha(s_{\alpha}^{(1)}+s_{\alpha}^{(2)}-pf_x)}{s_{\alpha}^{(1)}-s_{\alpha}^{(2)}}.
\end{equation}
Furthermore $Q,\tilde{Q}$ satisfy the equation for infinitesimal symmetries of ACH
\begin{equation}
Q_t=2pf_xQ+p^2\tilde{Q}_x.
\end{equation}

To generate the symmetries we apply the same idea as in the previous section. In this case 
we look at the asymptotic series solution of (\ref{tr1},\ref{tr2}) for small $|\alpha|$. This takes 
the form 
\[
s_{\alpha}=\sum_{n=1}^{\infty}s_n\alpha^{\frac{n}{2}},
\]
where
\[
s_1=p,~~~s_2=-\frac{p_x}{2},~~~s_{n+1}=-\frac{s_{n,x}}{2}+\frac{1}{2p}\left(\delta_{n,2}-\sum_{i=0}^{n-2}s_{i+2}s_{n-i}\right),~~~n=2,3...\, .
\]
The second solution of (\ref{tr1},\ref{tr2}) can be obtained by replacing $\alpha^{\frac{1}{2}}$ by $-\alpha^{\frac{1}{2}}$. Thus we take
\[
s_{\alpha}^{(1)}=\sum_{n=1}^{\infty}s_n\alpha^{\frac{n}{2}},~~~s_{\alpha}^{(2)}=\sum_{n=1}^{\infty}s_n(-\alpha^{\frac{1}{2}})^n.
\]
Plugging these into (\ref{symACH}) we obtain
\begin{equation}
\frac{Q(\alpha)}{\sqrt{\alpha}}=\frac{p\sum_{n=1}^{\infty}s_{2n}\alpha^{n}}{\sum_{n=1}^{\infty}s_{2n-1}\alpha^{n}}.\label{csch}
\end{equation}
The expansion of (\ref{csch}) around $\alpha=0$ gives an infinite hierarchy of symmetries of ACH.
The first few take the form
\begin{eqnarray}
X_{1}&=&p_x\frac{\partial}{\partial p}\ , \label{ACHX1}\\
X_{2}&=&\left(2p_{xxx}-12\frac{p_x}{p^2}-6\frac{p_xp_{xx}}{p}+3\frac{p_x^3}{p^2}\right)\frac{\partial}{\partial p}\, ,\label{ACHX2}\\
X_{3}&=&\left(\frac{2}{5}p_{xxxxx}-2\frac{p_xp_{xxxx}+2p_{xx}p_{xxx}}{p}+\frac{11p_xp_{xx}^2+7p_{x}^2p_{xxx}-4p_{xxx}}{p^2}\right.\nonumber\\
&+&\left.\frac{p_xp_{xx}(28-19p_x^2)}{p^3}+\frac{3p_x(16-40p_x^2+9p_x^4)}{4p^4}\right)\frac{\partial}{\partial p}.\label{ACHX3}
\end{eqnarray}
As far as we are aware these symmetries are new. As in the KdV case, it is straightforward to 
verify nontriviality. 

\subsection{SG}
The Sine Gordon equation 
\[
u_{xy}=\sin(u)
\]
can be brought to rational form by the change of variable $u=2i\ln(z)$, giving
\begin{equation}
zz_{xy}-z_xz_y=\frac{1}{4}(z^4-1).\label{SG}
\end{equation}
The BT for (\ref{SG}) \cite{Ba1} is $z\rightarrow z_{\alpha}$ where $z_{\alpha}$ satisfies
\begin{equation}
(z_{\alpha}z)_x=\frac{\alpha}{2}\left(z^2_{\alpha}-z^2\right),~~~zz_{\alpha,y}-z_yz_{\alpha}=\frac{1}{2\alpha}(z^2z_{\alpha}^2-1).\label{BTSG}
\end{equation}
The corresponding superposition principle is \cite{Hi3}
\begin{equation}
z_{\alpha,\beta}=z-(\alpha+\beta)\frac{z(z_{\alpha}-z_{\beta})}{\beta z_{\alpha}-\alpha z_{\beta}}\label{spsg}
\end{equation}
Plugging $\beta=-\alpha+\epsilon$ and expanding around $\epsilon=0$ we obtain
\begin{equation}
z_{\alpha,-\alpha+\epsilon}=z+\epsilon\frac{z(z_{\alpha}^{(1)}+z_{\alpha}^{(2)})}{\alpha( z_{\alpha}^{(1)}-z_{\alpha}^{(2)})}+O(\epsilon^2),\label{exSG}
\end{equation}
where $z_{\alpha}^{(1)}=z_{\alpha}$ and $z_{\alpha}^{(2)}=-z_{-\alpha}$ are distinct solutions of (\ref{BTSG}) for the same parameter value $\alpha$.
Relation (\ref{exSG}) describes an infinitesimal continuous transformation of $z$ which is also
a solution of SG. We denote it  $X(\alpha) = Q(\alpha) \frac{\partial}{\partial z}$ where
\begin{equation}
Q(\alpha)=\frac{z(z_{\alpha}^{(1)}+z_{\alpha}^{(2)})}{( z_{\alpha}^{(1)}-z_{\alpha}^{(2)})}.\label{symSG}
\end{equation}
One can check directly that $Q$ satisfies the equation for infinitesimal symmetries of SG
\[
Qz_{xy}+zQ_{xy}-Q_xz_y-z_xQ_y-z^3Q=0.
\]
There are two asymptotic solutions  of (\ref{BTSG})  for large $|\alpha|$, with form 
\[
z_{\alpha}=\sum_{n=0}^{\infty}\frac{v_n}{\alpha^n},
\]
where
\[
v_0=\pm z,~~~v_1= 2z_x,~~~ 
v_{n+1}=\frac{z_xv_n+zv_{n,x}-\frac{1}{2}\sum_{i=1}^{n}v_iv_{n+1-i}}{v_0},~~~n=2,3....
\]
Using these as the two different solutions $z_{\alpha}^{(1)}$ and $z_{\alpha}^{(2)}$ 
in (\ref{symSG}) and expanding in powers of $\frac1{\alpha}$ 
gives an infinite hierarchy of symmetries starting 
\begin{eqnarray*}
X_{1}&=&z_x\frac{\partial}{\partial z}\ , \\
X_{2}&=&\left(z_{xxx}-3\frac{z_xz_{xx}}{z}\right)\frac{\partial}{\partial z},\\
X_{3}&=&\left(z_{xxxxx}+\frac{20z_xz_{xx}^2}{z^2}+\frac{10z_x^2z_{xxx}}{z^2}-\frac{10z_{xx}z_{xxx}}{z}-\frac{10z_x^3z_{xx}}{z^3}-\frac{5z_xz_{xxxx}}{z}\right)\frac{\partial}{\partial z}.
\end{eqnarray*}
These symmetries coincide with ones obtained in \cite{Ol1}. However, as far as we know, 
the derivation above is new. 

\section{Algebra of the symmetries}
In this section we determine the algebra of the symmetries obtained in the previous section.  
We prove that symmetries generated by the Gardner method for KdV,
CH and SG commute. In the previous section we showed that symmetries for KdV can be obtained by 
the expansion of the generating symmetry $X(\alpha)$ in a series
around $\alpha=\infty$. In order to prove that these symmetries commute it is enough
to show the generating symmetry commutes with itself but with a different parameter,
namely that
\begin{equation}
[X(\alpha),X(\beta)]=0\ .\label{com1}
\end{equation}
(We refer in this section only to the symmetry for pKdV, the result follows
for the generating symmetry of KdV $Y(\alpha)$ by a simple prolongation argument.)
Here
\begin{equation}
X(\alpha) =\frac{1}{v_{\alpha}^{(1)}-v_{\alpha}^{(2)}} \frac{\partial}{\partial u}\ , ~~~
X(\beta)  =\frac{1}{v_{\beta}^{(1)}-v_{\beta}^{(2)}}   \frac{\partial}{\partial u} \ .
\end{equation}
To compute the commutator we need to know how $X(\alpha)$ acts on $v_{\beta}^{(1)}$
and $v_{\beta}^{(2)}$ and how $X(\beta)$ acts on $v_{\alpha}^{(1)}$ and $v_{\alpha}^{(2)}$.
This is currently not clear.
However we can also write $X(\alpha),X(\beta)$ in the form
\begin{equation}
X(\alpha) =\frac{1}{u_{\alpha}^{(1)}-u_{\alpha}^{(2)}} \frac{\partial}{\partial u}\ , ~~~
X(\beta)  =\frac{1}{u_{\beta}^{(1)}-u_{\beta}^{(2)}}   \frac{\partial}{\partial u} \ .
\end{equation}
The meaning of the first of these is that $X(\alpha)$ acts infinitesimally on $u$ via
$$ u\rightarrow u + \frac{\epsilon}{u_{\alpha}^{(1)}-u_{\alpha}^{(2)}}\ , $$
where we recall that $u_{\alpha}^{(1)},u_{\alpha}^{(2)}$ denote two distinct solutions of KdV obtained 
from $u$ by a B\"acklund transformation with parameter $\alpha$. Thus the action of $X(\alpha)$ 
on $u_\beta$ is 
$$ u_\beta\rightarrow u_\beta + \frac{\epsilon}{u_{\beta,\alpha}^{(1)}-u_{\beta, \alpha}^{(2)}}\ , $$
where $u_{\beta,\alpha}^{(1)},u_{\beta,\alpha}^{(2)}$ denote two distinct solutions of KdV obtained 
from $u_\beta$ by a B\"acklund transformation with parameter $\alpha$, which are 
of course determined by the superposition principle (\ref{eq2}).  
Thus the relevant prolongations of $X(\alpha)$ and $X(\beta)$ for computing the commutator are 
\begin{align}
\widehat{X}(\alpha)&=
 \frac{1}{u_{\alpha}^{(1)}-u_{\alpha}^{(2)}}\frac{\partial}{\partial u}
+\frac{1}{u_{\alpha,\beta}^{(1,1)}-u_{\alpha,\beta}^{(2,1)}}\frac{\partial}{\partial u_{\beta}^{(1)}}
+\frac{1}{u_{\alpha,\beta}^{(1,2)}-u_{\alpha,\beta}^{(2,2)}}\frac{\partial}{\partial u_{\beta}^{(2)}} \ , \\
\widehat{X}(\beta) &=
\frac{1}{u_{\beta}^{(1)}-u_{\beta}^{(2)}}\frac{\partial}{\partial u}
+\frac{1}{u_{\alpha,\beta}^{(1,1)}-u_{\alpha,\beta}^{(1,2)}}\frac{\partial}{\partial u_{\alpha}^{(1)}}
+\frac{1}{u_{\alpha,\beta}^{(2,1)}-u_{\alpha,\beta}^{(2,2)}}\frac{\partial}{\partial u_{\alpha}^{(2)}}\ .
\end{align}
To determine the commutator of $X(\alpha)$ and $X(\beta)$ we compute
\begin{eqnarray}
\widehat{X}(\alpha)Q(\beta) - \widehat{X}(\beta)Q(\alpha)
&=&
  \widehat{X}(\alpha)\left(\frac{1}{u_{\beta}^{(1)}-u_{\beta}^{(2)}}\right)
- \widehat{X}(\beta) \left(\frac{1}{u_{\alpha}^{(1)}-u_{\alpha}^{(2)}}\right)    \nonumber
\\
&=&
     \frac{1}{(u_{\beta}^{(1)}-u_{\beta}^{(2)})^2}
\left( -\frac{1}{u_{\alpha,\beta}^{(1,1)}-u_{\alpha,\beta}^{(2,1)}}
       +\frac{1}{u_{\alpha,\beta}^{(1,2)}-u_{\alpha,\beta}^{(2,2)}} \right) \label{eq70}
\\
&& -\frac{1}{(u_{\alpha}^{(1)}-u_{\alpha}^{(2)})^2}
\left( -\frac{1}{u_{\alpha,\beta}^{(1,1)}-u_{\alpha,\beta}^{(1,2)}}
       +\frac{1}{u_{\alpha,\beta}^{(2,1)}-u_{\alpha,\beta}^{(2,2)}}\right)\ .
\nonumber
\end{eqnarray}
In order to simplify this expression we have to use the superposition principle (\ref{eq2}).
Since we have two different BTs of $u$ for parameter $\alpha$ and two different BTs
of $u$ for parameter $\beta$ we need this in the four different forms
\begin{equation}
u^{(i,j)}_{\alpha,\beta}=u+\frac{\alpha-\beta}{u_{\alpha}^{(i)}-u_{\beta}^{(j)}},
~~~i,j\in\{1,2\}.\label{eq60}
\end{equation}
With the help of (\ref{eq60}) the right hand side of (\ref{eq70}) simplifies to zero,
thereby establishing (\ref{com1}). Thus  the symmetries obtained by the
Gardner method for pKdV (and KdV) generate an abelian group.

Let us prove a similar result for ACH. For this we need to compute the prolongation of the infinitesimal generator of the generating symmetry.
This can be done by the change of variable
\begin{equation}
u=\int p dx.\label{cv1}
\end{equation} The superposition principle for ACH can be rewritten in the form
\begin{equation}
u_{\alpha,\beta}=u-\left(\frac{(\alpha-\beta)(\alpha\beta-(u-u_{\alpha})(u-u_{\beta}))}{\beta (u-u_{\alpha})-\alpha (u-u_{\beta})}\right).\label{sp10}
\end{equation}
This expression is equivalent to the quad-graph equation 
$Q1_{\delta=1}$ in the ABS classification \cite{ABS}.
Using (\ref{cv1}), the generating symmetry of ACH is
\[
X(\alpha)=\frac{\alpha^2-(u-u_{\alpha}^{(1)})(u-u_{\alpha}^{(2)})}{u_{\alpha}^{(2)}-u_{\alpha}^{(1)}}\frac{\partial}{\partial u}.
\]
In order to prove that the symmetries commute it is enough
to show that (\ref{com1}) holds for the generating symmetry of ACH.
The relevant prolongations of $X(\alpha)$ and $X(\beta)$ are
\begin{align*}
\widehat{X}(\alpha)&=\frac{\alpha^2-(u-u_{\alpha}^{(1)})(u-u_{\alpha}^{(2)})}{u_{\alpha}^{(2)}-u_{\alpha}^{(1)}}\frac{\partial}{\partial u}+\frac{\alpha^2-(u_{\beta}^{(1)}-u_{\alpha,\beta}^{(1,1)})(u_{\beta}^{(1)}-u_{\alpha,\beta}^{(2,1)})}{u_{\alpha,\beta}^{(2,1)}-u_{\alpha,\beta}^{(1,1)}}\frac{\partial}{\partial u_{\beta}^{(1)}}\\&+\frac{\alpha^2-(u_{\beta}^{(2)}-u_{\alpha,\beta}^{(1,2)})(u_{\beta}^{(2)}-u_{\alpha,\beta}^{(2,2)})}{u_{\alpha,\beta}^{(2,2)}-u_{\alpha,\beta}^{(1,2)}}\frac{\partial}{\partial u_{\beta}^{(2)}} \ , \\
\widehat{X}(\beta) &=\frac{\beta^2-(u-u_{\beta}^{(1)})(u-u_{\beta}^{(2)})}{u_{\beta}^{(2)}-u_{\beta}^{(1)}}\frac{\partial}{\partial u}+\frac{\beta^2-(u_{\alpha}^{(1)}-u_{\alpha,\beta}^{(1,1)})(u_{\alpha}^{(1)}-u_{\alpha,\beta}^{(1,2)})}{u_{\alpha,\beta}^{(1,2)}-u_{\alpha,\beta}^{(1,1)}}\frac{\partial}{\partial u_{\alpha}^{(1)}}\\&+\frac{\beta^2-(u_{\alpha}^{(2)}-u_{\alpha,\beta}^{(2,1)})(u_{\alpha}^{(2)}-u_{\alpha,\beta}^{(2,2)})}{u_{\alpha,\beta}^{(2,2)}-u_{\alpha,\beta}^{(2,1)}}\frac{\partial}{\partial u_{\alpha}^{(2)}}\ .
\end{align*}
Relation (\ref{sp10}) can be presented in four different forms which connect different BTs for CH
\begin{equation}
u^{(i,j)}_{\alpha,\beta}=u-\left(\frac{(\alpha-\beta)(\alpha\beta-(u-u_{\alpha}^{(i)})(u-u_{\beta}^{(j)}))}{\beta (u-u_{\alpha}^{(i)})-\alpha (u-u_{\beta}^{(j)})}\right),~~~i,j\in\{1,2\}.
\end{equation}
With the help of this we obtain that (\ref{com1}) is true for the generating symmetry of ACH. Thus the symmetries obtained by the
Gardner method for ACH generate an abelian group.

The proof that symmetries obtained by the
Gardner method for SG commute is very similar to the above. The prolongations of the generating symmetries $X(\alpha)$ and $X(\beta)$ for SG
are
\begin{align*}
\widehat{X}(\alpha)&=\frac{z(z_{\alpha}^{(1)}+z_{\alpha}^{(2)})}{( z_{\alpha}^{(1)}-z_{\alpha}^{(2)})}\frac{\partial}{\partial z}+\frac{z_{\beta}^{(1)}(z_{\alpha,\beta}^{(1,1)}+z_{\alpha,\beta}^{(2,1)})}{( z_{\alpha,\beta}^{(1,1)}-z_{\alpha,\beta}^{(2,1)})}\frac{\partial}{\partial z_{\beta}^{(1)}}+\frac{z_{\beta}^{(2)}(z_{\alpha,\beta}^{(1,2)}+z_{\alpha,\beta}^{(2,2)})}{( z_{\alpha,\beta}^{(1,2)}-z_{\alpha,\beta}^{(2,2)})}\frac{\partial}{\partial z_{\beta}^{(2)}}\ , \\
\widehat{X}(\beta) &=\frac{z(z_{\beta}^{(1)}+z_{\beta}^{(2)})}{( z_{\beta}^{(1)}-z_{\beta}^{(2)})}\frac{\partial}{\partial z}+\frac{z_{\alpha}^{(1)}(z_{\alpha,\beta}^{(1,1)}+z_{\alpha,\beta}^{(1,2)})}{( z_{\alpha,\beta}^{(1,1)}-z_{\alpha,\beta}^{(1,2)})}\frac{\partial}{\partial z_{\alpha}^{(1)}}+\frac{z_{\alpha}^{(2)}(z_{\alpha,\beta}^{(2,1)}+z_{\alpha,\beta}^{(2,2)})}{( z_{\alpha,\beta}^{(2,1)}-z_{\alpha,\beta}^{(2,2)})}\frac{\partial}{\partial z_{\alpha}^{(2)}}\ .
\end{align*}
Relations which connect different BTs for SG can be obtained from (\ref{spsg})
\begin{equation}
z^{(i,j)}_{\alpha,\beta}=z-(\alpha+\beta)\frac{z(z_{\alpha}^{(i)}-z_{\beta}^{(j)})}{\beta z_{\alpha}^{(i)}-\alpha z_{\beta}^{(j)}},~~~i,j\in\{1,2\}.
\end{equation}
This enables us to prove (\ref{com1}) in the case of SG. Thus the symmetries obtained by the
Gardner method for SG also generate an abelian group.

\section{Connection of the Gardner method and Lenard recursion}
As mentioned in the introduction, there are several methods for generating symmetries of PDEs. 
One of them is Lenard recursion \cite{Ol1,Ol0,PS0}. In this section we show that Lenard recursion 
gives results equivalent to the Gardner method. We start with a brief explanation of Lenard 
recursion for symmetries of the KdV equation, and then show how this is related to the 
Gardner method. We then present similar results for the
cases of CH and SG. Finally, we  discuss the relation of Lenard recursion 
and the Gardner method for conservation laws in the case of KdV. 

Lenard recursion is based on the fact that KdV (\ref{eq}) is a bi-Hamiltonian system. 
Namely KdV can be presented in two forms:
\begin{align}
u_t&=P_0\frac{\delta H_0}{\delta\phi},\label{LR0}\\
u_t&=P_1\frac{\delta H_1}{\delta\phi},\label{LR1}
\end{align}
where 
\[H_0=\int\left(\frac{\phi^3}{2}-\frac{\phi_x^2}{8}\right)dx,~~~H_1=\int\frac{\phi^2}2dx,\]
and
\[P_0=\frac{\partial}{\partial x},~~~P_1=\frac{1}{4}\frac{\partial^3}{\partial x^3}+2\phi\frac{\partial}{\partial x}+\phi_x. \]
The expression $\frac{\delta}{\delta\phi}$ denotes the variational derivative.
If $g=\int Gdx$ is a conserved quantity for KdV equation then both $Q_0=P_0\frac{\delta g}{\delta\phi}, Q_1=P_1\frac{\delta g}{\delta\phi}$ are the characteristics of symmetries.
Evidently
\[
P_1P_0^{-1}Q_0=Q_1.
\]
Thus the ``recursion operator for symmetries''
\[
R=P_1P_0^{-1}
\]
maps Hamiltonian symmetries to Hamiltonian symmetries, and can be used to generate 
the infinite hierarchy of symmetries of KdV. However it is necessary to check that at 
each step the operator $P_0^{-1}$ can be applied.

In section 2 we showed that KdV has a symmetry with characteristic
\[Q(\alpha)_x=\left(\frac{1}{v_{\alpha}^{(1)}-v_{\alpha}^{(2)}}\right)_x=\frac{v_{\alpha}^{(1)}+v_{\alpha}^{(2)}}{v_{\alpha}^{(1)}-v_{\alpha}^{(2)}}.\]
Applying $R$ to $Q(\alpha)_x$ we obtain
\begin{equation}
RQ(\alpha)_x=P_1P_0^{-1}\left(\frac{1}{v_{\alpha}^{(1)}-v_{\alpha}^{(2)}}\right)_x=P_1\frac{1}{v_{\alpha}^{(1)}-v_{\alpha}^{(2)}}=\frac{\alpha(v_{\alpha}^{(1)}+v_{\alpha}^{(2)})}{v_{\alpha}^{(1)}-v_{\alpha}^{(2)}}=\alpha Q(\alpha)_x.\label{RO1}
\end{equation}
Writing
\[Q(\alpha)_x=\frac{1}{\alpha^{1/2}}\sum_{i=0}^{\infty}\frac{Q_i}{\alpha^{i}}\]
and substituting  into (\ref{RO1}) we obtain
\[R\left(\sum_{i=0}^{\infty}\frac{Q_i}{\alpha^{i}}\right)=\sum_{i=0}^{\infty}\frac{Q_i}{\alpha^{i-1}}.\]
Comparing coefficients of powers of $\alpha$ we get
\[
RQ_i=Q_{i+1},~~~i=0,1,2,....
\]
Thus Lenard recursion gives the same symmetries as the Gardner method for KdV.
Furthermore, in our approach we see immediately that all the characteristics are $x$ derivatives.
Also note that application of the recursion operator $R$ on $Q(\alpha)$ is equivalent 
to multiplication by the parameter $\alpha$.  

Similar results can be reproduced for ACH and SG, and we briefly give these here. 
For ACH we recall the generating symmetry is given by (\ref{symACH}). 
It is straightforward to check, using (\ref{tr2}), that 
\[
Q= \partial_t \left( \frac{p}{s_{\alpha}^{(1)}-s_{\alpha}^{(2)}} \right) ,
\]
and we consequently identify $\partial_t^{-1} Q$ with the quantity on parentheses on the RHS. 
Then a simple calculation using (\ref{tr1}) gives 
\begin{equation}
R Q = \frac1{\alpha} Q 
\end{equation}
where $R$ is the recursion operator for ACH, 
\[
R=- p_x\partial_t^{-1} + \frac14\left( 
\partial_x^{2} -\frac{p_x}{p}\partial_{x} +\frac{p_{x}^2-pp_{xx}-4}{p^2}
\right).
\]
Note that in the case of CH (as opposed to KdV and SG) the expansion of the generating 
symmetry is in increasing (as opposed to decreasing) powers of $\alpha$. The recursion
operator can be used directly to generate the symmetries, but a double miracle
is needed at each step: The existing symmetry has to be a derivative with respect
to $t$, and there is no apparent guarantee that the new symmetry will turn out to 
depend only on $p$ and its $x$-derivatives. For example, for application of the recursion 
operator to the symmetry $X_1$ given in (\ref{ACHX1}) the following equality is necessary:
\[
p_x = \partial_t \left( \frac{1}{2p^2}+\frac{p_x^2}{8p^2}+\frac{1}{4}(\ln p)_{xx} \right).
\]
Application of the operator to higher symmetries is apparently only more difficult.

For SG, starting with equation (\ref{symSG}) and using the first of equations 
(\ref{BTSG}) to differentiate $z_\alpha^{(1)},z_\alpha^{(2)}$ with respect to $x$, it is 
easy to check that 
\begin{eqnarray}
\left(\frac{Q}{z} \right)_x &=&  
- \frac{\alpha(  z_\alpha^{(1)} z_\alpha^{(2)} + z^2 )} 
       {z ( z_\alpha^{(1)} -  z_\alpha^{(2)} )  } \\
\left(\frac{Q}{z} \right)_{xx} - \alpha^2 \left(\frac{Q}{z}\right) &=&  
  \frac{2\alpha z_x ( z_\alpha^{(1)} z_\alpha^{(2)} - z^2) }
       {z^2(z_\alpha^{(1)} - z_\alpha^{(2)}) }  \label{j2}
\end{eqnarray}
Furthermore  
\begin{equation}
\partial_x \left( \frac{\alpha(z_\alpha^{(1)}z_\alpha^{(2)}-z^2)}
                {2z(z_\alpha^{(1)}-z_\alpha^{(2)})}  \right) 
= - \frac{\alpha z_x(  z_\alpha^{(1)} z_\alpha^{(2)} + z^2 )} 
       {z^2 ( z_\alpha^{(1)} -  z_\alpha^{(2)} )  } 
=\frac{z_x}{z} \left(\frac{Q}{z}\right)_x  \label{j3}
\end{equation} 
Combining (\ref{j2}) and (\ref{j3}) gives 
\begin{equation}
\left(\frac{Q}{z} \right)_{xx} - \alpha^2 \left(\frac{Q}{z}\right) =
\frac{4z_x}{z} \partial_x^{-1} 
\left( \frac{z_x}{z} \left(\frac{Q}{z}\right)_x  \right) 
\label{j4}\end{equation} 
where the precise meaning of the term $\partial_x^{-1}(\ldots)$ on 
the right hand side of (\ref{j4}) is the term in 
parentheses in the left hand side of (\ref{j3}). To interpret 
(\ref{j4}) we return to the standard Sine-Gordon variable $u=2i\ln z$ 
and write $Q'=\frac{Q}{z}$ to write (\ref{j4}) in the form 
\begin{equation}
Q'_{xx}  + u_x \partial_x^{-1} ( u_x Q'_x) = \alpha^2  Q' 
\label{j5}\end{equation} 
$Q'$ is the generating symmetry of Sine-Gordon in the $u$ variable. 
Its $\alpha$ expansion takes the form 
\begin{equation}
Q' = \sum_{n=0}^\infty \frac{C_n}{\alpha^{2n+1}}, 
\end{equation} 
where $C_0=u_x$ (up to a rescaling) and, from (\ref{j5}), 
\begin{equation}
C_{n+1} = \left( \partial_{xx}  + u_x \partial_x^{-1} u_x \partial_x  
    \right) C_n \ .   
\label{j7}\end{equation} 
The operator on the RHS of (\ref{j7}) is precisely the recursion 
operator for symmetries given by Olver \cite{Ol1}.

Although the focus of this paper is symmetries,
we conclude this section with a discussion of the relation of Lenard recursion 
and the Gardner method for conservation laws in the case of KdV, as 
we found the necessary calculations rather tricky. In the Gardner method the 
infinite hierarchy of conservation laws for KdV is obtained by expanding the 
conservation law
\begin{equation}
\partial_t (v_{\alpha}^{(1)}-v_{\alpha}^{(2)}) + \partial_x \left(  - (u+\alpha)(v_{\alpha}^{(1)}-v_{\alpha}^{(2)}) \right)
=0.\label{CL100}\end{equation}
into a series in $\alpha$ with the help of (\ref{exps}) \cite{Miu1}.
The conserved quantity associated with (\ref{CL100}) has the form
$g=\int (v_{\alpha}^{(1)}-v_{\alpha}^{(2)})dx$.
We now prove that the variational derivative of $g$ is
\begin{equation}
\frac{\delta g}{\delta\phi}=\frac{4}{v_{\alpha}^{(1)}-v_{\alpha}^{(2)}}.\label{LR10}
\end{equation}
The proof for (\ref{LR10}) is as follows. We know from (\ref{BT10}) that
\[
v^{(1)}_{\alpha,x}+v^{(2)}_{\alpha,x}-2\alpha+(v_{\alpha}^{(1)})^2+(v_{\alpha}^{(2)})^2=-4\phi
\]
The variation of this expression is
\[
\delta v^{(1)}_{\alpha,x}+\delta v^{(2)}_{\alpha,x}+2v_{\alpha}^{(1)}\delta v_{\alpha}^{(1)}+2v_{\alpha}^{(2)}\delta v_{\alpha}^{(2)}=-4\delta\phi.
\]
Dividing this by $v_{\alpha}^{(1)}-v_{\alpha}^{(2)}$ we obtain
\[
\frac{\delta v^{(1)}_{\alpha,x}+\delta v^{(2)}_{\alpha,x}}{v_{\alpha}^{(1)}-v_{\alpha}^{(2)}}+\frac{2v_{\alpha}^{(1)}\delta v_{\alpha}^{(1)}+2v_{\alpha}^{(2)}\delta v_{\alpha}^{(2)}}{v_{\alpha}^{(1)}-v_{\alpha}^{(2)}}=-4\frac{\delta\phi}{v_{\alpha}^{(1)}-v_{\alpha}^{(2)}}.
\]
This expression can be rewritten differently as
\[
-\frac{(\delta v_{\alpha}^{(1)}+\delta v_{\alpha}^{(2)})(v_{\alpha}^{(1)}+v_{\alpha}^{(2)})}{v_{\alpha}^{(1)}-v_{\alpha}^{(2)}}+\partial_x\left(\frac{\delta v_{\alpha}^{(1)}+\delta v_{\alpha}^{(2)}}{v_{\alpha}^{(1)}-v_{\alpha}^{(2)}}\right)+\frac{2v_{\alpha}^{(1)}\delta v_{\alpha}^{(1)}+2v_{\alpha}^{(2)}\delta v_{\alpha}^{(2)}}{v_{\alpha}^{(1)}-v_{\alpha}^{(2)}}=-4\frac{\delta\phi}{v_{\alpha}^{(1)}-v_{\alpha}^{(2)}}.
\]
After simplification we obtain
\[\delta v_{\alpha}^{(1)}-\delta v_{\alpha}^{(2)}=4\frac{\delta\phi}{v_{\alpha}^{(1)}-v_{\alpha}^{(2)}}+\partial_x\left(\frac{\delta v_{\alpha}^{(1)}+\delta v_{\alpha}^{(2)}}{v_{\alpha}^{(1)}-v_{\alpha}^{(2)}}\right).
\]
Thus
\[
\delta\int(v_{\alpha}^{(1)}-v_{\alpha}^{(2)})dx=\int\frac{4}{v_{\alpha}^{(1)}-v_{\alpha}^{(2)}}\delta\phi dx,
\]
and (\ref{LR10}) is proved.
We have already seen in (\ref{RO1}) that
\[
P_1\frac{1}{v_{\alpha}^{(1)}-v_{\alpha}^{(2)}}=\alpha Q(\alpha)_x=\alpha P_0\frac{1}{v_{\alpha}^{(1)}-v_{\alpha}^{(2)}}.
\]
Thus
\[P_1\frac{\delta g}{\delta\phi}=\alpha P_0\frac{\delta g}{\delta\phi},
\]
or
\[
\bar{R}\frac{\delta g}{\delta\phi}=\alpha \frac{\delta g}{\delta\phi},
\]
where $\bar{R}=P_0^{-1}P_1$ is the recursion operator for (variational derivatives of) conserved quantities of KdV.

\color{black}
\section{Connection with symmetries of difference equations.}
Relation (\ref{eq2})
\[
u_{\alpha,\beta}=u+\frac{\alpha-\beta}{u_{\alpha}-u_{\beta}}
\]
connects solutions of the pKdV equation. It can be embedded on the quad-graph,
the planar graph with quadrilateral faces \cite{ABS,AV0,NC0,NQC,Je0}.
The embedded equation is called the discrete Korteweg-de Vries equation (dKdV)
and has the following form
\begin{equation}
u_{1,1}=u_{0,0}+\frac{\alpha-\beta}{u_{1,0}-u_{0,1}}\label{qg1}.
\end{equation}
Here $k,l\in \mathbb{Z}^2$ are independent variables
and $u_{0,0}=u(k,l)$ is a dependent variable that is defined on the domain
$\mathbb{Z}^2$. We denote the values of this variable on
other points by $u_{i,j}=u(k+i,l+j)=S_k^iS_l^ju_{0,0}$, where $S_k,~S_l$ are
unit forward shift operators in $k$ and $l$ respectively.
Let us explain the connection between (\ref{qg1}) and (\ref{eq2}).
Shifts $u_{1,0}$ and $u_{0,1}$ are BTs of $u=u_{0,0}$ for two different parameters
$\alpha, \beta$.
In general $u_{i,j}$ denotes the application of the BT $i+j$ times to $u=u_{0,0}$,
$i$ times with parameter $\alpha$ and $j$ times with parameter $\beta$.

Properties of dKdV are closely related to properties of pKdV. In particular,
since the continuous generating symmetry of pKdV discussed in the previous sections
is an infinitesimal version of a B\"acklund transformation, and B\"acklund
transformations commute, the generating symmetry must also give a continuous
symmetry of dKdV, and we wish to identify its generator.

So far we have identified the functions $u_{\alpha}^{(1)},u_{\alpha}^{(2)}$ in the
pKdV generating symmetry
\begin{equation}
X(\alpha)=\frac{1}{u_{\alpha}^{(1)}-u_{\alpha}^{(2)}}\frac{\partial}{\partial u}
\end{equation}
as different B\"acklund transformations of $u$ with the same parameter $\alpha$.
In fact it is straightforward to show that if the solution $u_\alpha$ of KdV is 
generated from $u$ by a BT with parameter $\alpha$, then the solution $u$ 
is generated from $u_\alpha$ by a BT with parameter $\alpha$. Thus we can 
identify $u_{\alpha}^{(1)}$, say, as a forward B\"acklund transformation of $u$
and $u_{\alpha}^{(2)}$ as a reverse B\"acklund transformation. 
In the language of the quad-graph, inverse shifts  $u_{-1,0}$ and $u_{0,-1}$ are
reverse BTs. So we can identify $u_{\alpha}^{(2)}$ with $u_{-1,0}$ and $u_{\alpha}^{(1)}$ with
$u_{1,0}$ and the quad-graph version of $X(\alpha)$ is thus
\begin{equation}
X_1=\frac{1}{u_{1,0}-u_{-1,0}}\frac{\partial}{\partial u_{0,0}} \ .
\end{equation}
Similarly the quad-graph version of $X(\beta)$ is
\begin{equation}
X_2=\frac{1}{u_{0,1}-u_{0,-1}}\frac{\partial}{\partial u_{0,0}} \ .
\end{equation}
These symmetries were already found in \cite{RH2}.
Furthermore, in \cite{RS1} it was shown that equation (\ref{qg1})
embedded in three dimensions has another symmetry of the form
\begin{equation}
X=\frac{1}{u_{0,0,1}-u_{0,0,-1}}\frac{\partial}{\partial u_{0,0,0}}\ .
\end{equation}
This evidently also has its origins in the generating symmetry.
In \cite{RS1} it was shown how, by a suitable expansion,
this symmetry gives the infinite hierarchy of symmetries for the dKdV equation.

As we noted before the CH equation after the change of variable $u=\int p dx$ is related to quad-graph equation $Q1_{\delta=1}$. Namely, the superposition principle for CH (\ref{sp10}) is equivalent to quad-graph equation $Q1_{\delta=1}$ in the ABS classification \cite{ABS}
\[
\theta_2(u_{1,1}-u_{0,1})(u_{0,0}-u_{1,0})-\theta_1(u_{1,1}-u_{1,0})(u_{0,0}-u_{0,1})+(\theta_1-\theta_2)\theta_1\theta_2=0.
\]
So from the generating symmetry of ACH we can obtain the symmetry of $Q1_{\delta=1}$.

By doing the same computation as we did for pKdV we can
identify $u_{\alpha}^{(1)}$ as a forward B\"acklund transformation
and $u_{\alpha}^{(2)}$ as a reverse B\"acklund transformation. Thus
\begin{align*}
X_1&=\left(\frac{\alpha^2-(u_{0,0}-u_{1,0})(u_{0,0}-u_{-1,0})}{u_{-1,0}-u_{1,0}}\right)\frac{\partial}{\partial u_{0,0}},\\
X_2&=\left(\frac{\alpha^2-(u_{0,0}-u_{0,1})(u_{0,0}-u_{0,-1})}{u_{0,-1}-u_{0,1}}\right)\frac{\partial}{\partial u_{0,0}}.
\end{align*}
are the symmetries of $Q1_{\delta=1}$.

For SG we can
identify $z_{\alpha}^{(1)}$ as a forward B\"acklund transformation and $-z_{\alpha}^{(2)}$ as a reverse B\"acklund transformation.
From generating symmetry of SG we obtain that
\begin{align*}
X_1&=\frac{z_{0,0}(z_{1,0}-z_{-1,0})}{( z_{1,0}+z_{-1,0})}\frac{\partial}{\partial z_{0,0}},\\
X_2&=\frac{z_{0,0}(z_{0,1}-z_{0,-1})}{( z_{0,1}+z_{0,-1})}\frac{\partial}{\partial z_{0,0}}.
\end{align*}
are the symmetries for equation
\[
\alpha(z_{0,0}z_{0,1}-z_{1,0}z_{1,1})-\beta(z_{0,0}z_{1,0}-z_{0,1}z_{1,1})=0.
\]
This equation is equivalent to $H3_{\delta=0}$ in the ABS classification.
\section{Conclusion}
In this paper we have introduced the Gardner method for generation of
the infinite hierarchy of symmetries of integrable equations, using
KdV, CH and SG as examples. The method involves identifying the
generating symmetry $X(\alpha)$ from the superposition principle for BTs
of the equations studied,  followed by a suitable expansion in powers
of $\alpha$. The method is both mathematically elegant and
computationally efficient. We have shown how to use our formalism to
prove the symmetries commute, explained the origin of Lenard recursion
relations, and explored the link with integrable lattice equations.
The fact that integrable lattice equations arise as the superposition
principle for BTs  of continuum equations is well-known --- indeed the
celebrated $Q4$ lattice equation in the ABS classification was first
written down by Adler \cite{Adl2} as the superposition principle for the
Krichever-Novikov equation. However, we believe the link between the
$Q1_{\delta=1}$ lattice equation and CH given in this paper is new, as
are the symmetries derived for the ACH equation.

We expect the method to be applicable to other integrable equations,
and it will be interesting to see more examples developed. It would be
particularly interesting to see examples in which infinite hierarchies
of symmetries and/or conservation laws can be constructed by one
method, but not by another (though this may be difficult to verify).
\vskip.3in

\noindent
{\bf \Large Acknowledgments}
We thank Peter Hydon for encouraging this line of research and Qiming Liu 
and the referees for significant comments.

\end{document}